\documentclass[useAMS,usenatbib]{mn2e}
\usepackage{graphicx}

\def\simlt{\lower.5ex\hbox{\simlt}}
\def\gtsima{$\; \buildrel > \over \sim \;$}
\def\simgt{\lower.5ex\hbox{\gtsima}}

\def\aj{\rm AJ}
\def\apj{\rm ApJ}
\def\aap{\rm A\&A}
\def\mnras{\rm MNRAS}
\def\araa{\rm ARA\&A}

\title[The Luminosity Function and M/L ratio of NGC~2419]{The Luminosity Function and stellar Mass to Light ratio of the massive globular cluster NGC~2419\thanks{Based on observations made with the NASA/ESA Hubble Space
Telescope, obtained at the Space Telescope Science Institute, which is operated by the Association of Universities for Research in Astronomy, Inc., under NASA contract NAS 5-26555. These observations are asso- ciated with program GO-11903 [P.I.: J. Kalirai].}}
\author[M. Bellazzini et al.]{M. Bellazzini$^{1}$\thanks{E-mail:
michele.bellazzini@oabo.inaf.it}, E. Dalessandro$^{2}$, A. Sollima$^{3}$, R. Ibata$^{4}$ 
\\ %\\
$^{1}$INAF-Osservatorio Astronomico di Bologna, Via Ranzani 1, 40127, Bologna, Italy.\\
$^{2}$Dip. di Astronomia - Univ. di Bologna, Via Ranzani 1, 40127, Bologna, Italy.\\
$^{3}$INAF-Osservatorio Astronomico di Padova, vicolo dell'Osservatorio 5, 35122, Padova, Italy.\\
$^{4}$Observatoire Astronomique, Universit\'e de Strasbourg, CNRS, 11, rue de l'Universite', F-67000 Strasbourg, France.\\
}
\begin{document}

\date{Accepted for publication in MNRAS, March 13, 2012}

\pagerange{\pageref{firstpage}--\pageref{lastpage}} \pubyear{2012}

\maketitle

\label{firstpage}

\begin{abstract}
We used archival Hubble Space Telescope WFC3 images to obtain the Luminosity Function of the remote globular cluster NGC~2419 from two magnitudes above the Horizontal Branch level down to $\simeq 3.0$ magnitudes below the Turn Off point (to $M_I\simeq 6.4$), approximately covering the range of initial stellar masses $ 0.5~M_{\sun}\la m\la 0.9~M_{\sun}$. The completeness-corrected Luminosity Function does not display any change of shape over the radial range covered by the WFC3 data, out to $\simeq 6$ core radii ($r_c$), or, equivalently, to $\simeq 2$ half-light radii.
The Luminosity Function in this radial range is also identical to that obtained from ground based data at much larger distances from the cluster centre ($12r_c\la R\la 22r_c$), in the magnitude range in which the two distributions overlap ($M_I\le 4.0$). These results support the conclusion by Dalessandro et al. that there is no significant mass segregation among cluster stars, hence the stellar mass-to-light ratio remains constant with distance from the cluster centre. We fitted the observed Luminosity Function with theoretical counterparts with the proper age and metallicity from different sets of stellar evolution models and we consistently derive a total V band mass-to-light ratio $1.2\la M/L_V\la 1.7$, by extrapolating to the Hydrogen burning limit, with a best-fit value $M/L_V=1.5 \pm 0.1$. On the other hand, assuming that there are no cluster stars with $m\le 0.3~M_{\sun}$, we establish a robust lower limit
$M/L_V> 0.8$. These estimates provide useful constraints for dynamical models of the cluster that were forced to consider the stellar mass-to-light ratio as a (nearly) free parameter.    
\end{abstract}


\section{Introduction}

The remote and massive globular cluster (GC) NGC~2419 is a curious stellar system in several aspects. For example, it is the only bright ($M_V=-9.5$) Galactic GC lying at Galactocentric distances larger than 30~kpc \citep[see Fig.~8, in][]{silvia07}. It has a half-light radius much larger than other GCs of the same luminosity \citep[][and references therein]{mack,mic}, more akin to stellar nuclei or Ultra Compact Dwarf galaxies \citep[UCD][]{evsti,brodie} than to classical globulars. It displays a very extended bi-modal Horizontal Branch \citep[HB][]{sand,ema}, possibly suggesting a significant spread in He abundance \citep{dic_mult}. A small spread in calcium and iron abundance has also been claimed, recently, based con Ca triplet spectroscopic surveys \citep{judylr,iba11a}. However the iron abundance of seven stars studied by \citet{judyhr} with high-resolution spectra appears remarkably homogeneous. \citet[][D08 hereafter]{ema} used deep photometry over a field covering the whole cluster to demonstrate that Blue Straggler stars \citep[BSS,][]{baibss} in NGC~2419 have the same radial distribution as Red Giant Branch (RGB) and HB stars (as well as the overall cluster light). The same behaviour has been observed only in the very peculiar massive cluster $\omega$~Cen \citep{omega} and in the sparse and remote cluster Pal~14 \citep{pal14}, while in all the other GCs, BSS are strongly centrally concentrated with respect to other stellar species, as a consequence of mass segregation driven by two-body relaxation \citep{bss}.

In recent times, NGC~2419 has been the object of several kinematic studies aimed at deriving the mass distribution within the cluster and at testing alternative theories of gravitation \citep{baum09,conroy,iba11a,iba11b}. This is due to several interesting (and in some case unique) properties of this system.
Its large Galactocentric distance ($R_{GC}\simeq 95$~kpc) implies that the effect of Galactic tidal forces on the cluster are negligible, so that the motion of their stars at any distance from the cluster centre can be safely interpreted as driven only by the cluster potential and the law of gravitation \citep{baum09,iba11a}. The total luminosity of NGC~2419 is $\sim 25$ times larger than any other Galactic GC lying in these remote regions. This means that the potential targets for estimating individual radial velocities, i.e. RGB stars reachable with medium resolution spectrographs on 8m class telescopes, are fairly abundant and can allow to sample adequately the velocity dispersion curve out to large distances \citep[see][]{iba11a}, while in other distant galactic GCs they are barely sufficient to derive a global value of the dispersion \citep[see, e.g.,][]{jordi,Apal14}. The gas-less nature and spherical symmetry typical of old globulars greatly simplifies the modelling of the system.

Since it is apparent that the cluster potential is dominated by the mass provided by stars, any independent constraint on the stellar mass-to-light ratio (M/L) and its radial behaviour can provide precious support to such studies. Since GCs stars are coeval and chemically homogeneous\footnote{In the considered case any small inhomogeneity \citep{judylr,judyhr,dic_mult} should have negligible effects on the stellar mass-to-light ratio.}, given a fixed mass-luminosity relation \citep[MLR; see, e.g.][and references therein]{paust09}, (a) the total stellar mass-to-light ratio depends only on the form of the present-day Mass Function (MF\footnote{Since we are interested in the {\em current} status of the cluster we consider only the present-day Mass Function, not the Initial Mass Function that is the subject of many studies in this field. In the following we refer to the present-day Mass Function simply as MF.}; except for the contribution of dark remnants, that is not dominant, see below), and (b) the radial variation of the stellar M/L depends only on the degree of mass segregation within the cluster. Both issues can be explored observationally by deriving a completeness-corrected Luminosity Function (LF) and looking for any radial variation of the LF.
This is the aim of the present analysis. We use the deepest images available for NGC~2419, obtained with the Wide Field Camera 3 (WFC3) on board of the Hubble Space Telescope (HST), to study the LF of the cluster in different radial ranges.

The techniques to obtain reliable luminosity (and mass) functions in GCs, corrected for all the observational effects, are well established and have been widely used for several years \citep[see][for recent applications, review and references]{paust09,paust10,guido,bastian}. While the MF of nearby clusters can be traced down to 0.10-0.15~$M_{\sun}$ (i.e. near the Hydrogen burning limit) with HST photometry reaching $I\sim 25.0$ \citep{paust10}, the large distance to NGC~2419 limits the accessible range of masses to $m\la 0.5~M_{\sun}$ even if our photometry reaches $I\simeq 27.5$ (see below). Therefore the portion of the cluster MF providing the largest contribution to the stellar mass budget\footnote{To provide a quantitative idea, from an isochrone of age=12 Gyr, [Fe/H]=-2.1 and [$\alpha$/Fe]+0.4 from the Dartmouth set \citep{dotter}, a MS star at the Hydrogen burning limit has $M/L\simeq 60$,
one with m=0.4~$M_{\sun}$ has $M/L\simeq 11$, and one at the Turn Off point $M/L\simeq 0.3$, while a giant at the RGB bump has $M/L\simeq 0.01$ and at the RGB Tip $M/L\simeq 0.0004$ (bolometric luminosity).} is out of reach of the present study. This will also limit our ability to detect radial variations, since a larger range of stellar masses sampled would increase the sensitivity to subtler effects. Moreover, the relatively small field covered by the WFC3 data also limits the explored radial range to $\simeq 2$ half-light radii ($r_h$).
Still, there are solid reasons to believe that the results that can be obtained from this analysis are worth the effort:

\begin{enumerate}

\item While the total V band mass-to-light ratio predicted by stellar evolution models are typically around $\simeq 2$ \citep{krui_ml}, direct dynamical estimates as low as $M/L_V=0.4-0.5$ are reported for some clusters \citep{mandu_ml,pm93,strader}. In some cases this can be due to observational limitations\footnote{For example, \citet{pm93} report $M/L_V=0.7$ at the centre of NGC~2419 based on an estimate of the central velocity dispersion $\sigma_0=3.0$~km~s$^{-1}$, derived from a sample of 12 stars presented by \citep{ols}. However, recent studies based on much larger samples, demonstrated that the actual value of $\sigma_0$ is clearly higher than 6 ~km~s$^{-1}$, implying $M/L_V\sim 2$ \citep{baum09,iba11a}.}, but, in general, it is attributed to preferential loss of low mass stars, due to the combined effects of mass-segregation and evaporation in a tidal field \citep[see][for discussion and references]{krui_ml}. Thus, even a best-effort estimate or a robust lower-limit based on the {\em real} MF would be valuable in narrowing the parameter space to be explored with dynamical models. 

\item Detailed theoretical analyses \citep[e.g.][]{bm01} show that the effects of mass-segregation should be quite evident in the inner regions of clusters and the slope of the MF is observed to change also on mass ranges smaller than that considered here. 

\item We use the deep and wide-field photometry by D08 to extend the analysis on the radial variation of the MF over the whole body of the cluster, albeit  limited to an even smaller range of sampled mass.

\end{enumerate}

%%%%%%%%%%%%%%%%%%%%%%%%%%%%%%%%%%%%%%%%%%%%%%%%%%%%%%%%%%%%%%%%%%%%%%%%% TABELLA PARAMETRI
\begin{table}
\label{tabpar}
 \centering
 \begin{minipage}{70mm}
  \caption{Adopted cluster parameters.}
  \begin{tabular}{@{}lcc@{}}
  \hline
   Parameter     &  value & Ref.\\
  \hline
 $(m-M)_0$   & 19.71$\pm$0.08 &\citet{dic_dist}\\
 $E(B-V)$    & 0.08$\pm$0.01   &\citet{dic_dist}\\
 $r_c$       & 27.6$\arcsec$  &\citet{iba11a}\footnote{For their best-fit model (\#17).}\\
 $r_h$       & 56.3$\arcsec$  &\citet{iba11a}$^a$\footnote{In excellent agreement with the indepedent estimate by \citet{mic}.}\\
 $V_t$       & 10.47$\pm$0.07 &\citet{mic}\\
 $M_V$       & -9.5$\pm$0.1\footnote{From the values listed above.}   &           \\
 $[Fe/H]$    & -2.1$\pm$0.1                                           & \citet{judyhr}\\
 $[\alpha/Fe]$& +0.4\footnote{Average of $[O/Fe]$, $[Mg/Fe]$, $[Si/Fe]$, and $[Ca/Fe]$.}& \citet{judyhr}\\
 $R_{F606W}$ \footnote{Where $R_{\lambda}=A_{\lambda}/E(B-V)$.}& 2.874 & \citet{marigo}\footnote{From  {\tt http://stev.oapd.inaf.it/cgi-bin/cmd} for the WFC3 VEGAMAG system}\\
 $R_{F814W}$~$^d$ & 1.894 & \citet{marigo}$^e$\\ \hline
\end{tabular}
\end{minipage}
\end{table}
%%%%%%%%%%%%%%%%%%%%%%%%%%%%%%%%%%%%%%%%%%%%%%%%%%%%%%%%%%%%%%%%%%%%%%%%% FINE TABELLA PARAMETRI

\begin{figure}
\includegraphics[width=80mm]{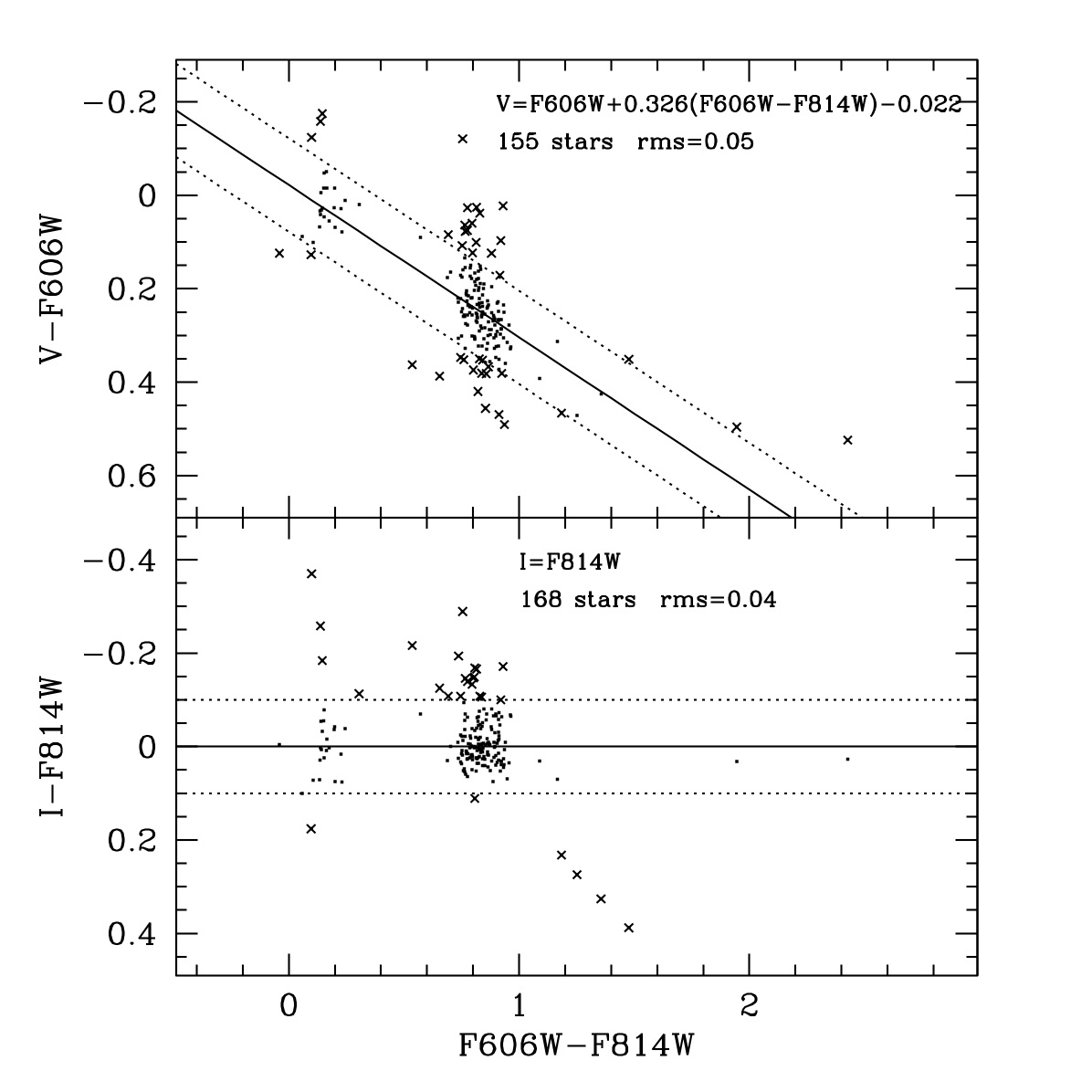}
 \caption{Transformations between the WFC3 VEGAMAG F606W and F814W magnitudes and the 
 Johnson-Kron-Cousins V and I magnitudes, obtained from stars in common between our WFC3 sample and the secondary photometric standard sample by \citet{s00}. Stars marked with a cross have been rejected, since they are outside the $\pm 0.1$ mag band (dotted lines) around the best fit (44 rejected stars in V-F606W and 31 in I-F814W).}
 \label{trabel}
\end{figure}

In the following we always adopt the cluster parameters listed in Table~\ref{tabpar}, that are the most reliable and up-to-date, from the literature. Since the MLR must be the same at all radii, comparing LFs in different radial ranges is strictly equivalent to comparing the MFs. Also, theoretical LFs can be produced for any assumed IMF\footnote{In all these models the IMF is then evolved into a LF/MF of the chosen age, thus allowing the comparison with the observed present-day LFs \citep[see][]{rf88}.} using web tools that will be described below. Hence we avoid the transformation of the observed LFs into the corresponding MFs since it is not needed in the present context. Throughout  the paper we will consider LFs in the F814W VEGAMAG HST-WFC3 passband. In addition to the lower sensitivity to interstellar reddening with respect to F606W, we found that F814W can be safely transformed into Cousins' I, thus allowing an easy comparison with ground-based data. In Fig.~\ref{trabel} we compare F606W, F814W magnitudes with Johnson-Kron-Cousins V,I, as a function of colour, for more than 150 stars in common between our WFC3 photometry of NGC~2419 and the set of secondary standards by \citet{s00,s05}. While the transformation between V and F606W display a strong (and possibly non linear) dependence on colour and a large scatter at any colour, I and F814W appear identical independently of colour, within the uncertainties. For this reason we will use F814W and I interchangeably in the following.
Note that the comparison of Fig.~\ref{trabel} is performed with stars lying within 120$\arcsec$ of the cluster centre, where crowding is expected to  provide significant contribution to the r.m.s. scatter. Therefore, it is likely that magnitudes in the two filters do coincide to within less than $\sim 0.02$~mag, at least for metal-poor stars in this range of colour.

%%%%%%%%%%%%%%%%%%%%%%%%%%%%%%%%%%%%%%%%%%%%%%%%%% FIG 
%%%%%%%%%%%%%%%%%%%%%%%%%%%%%%%%%%%%%%%%%%%%%%%%%%%%%%%%%%%%%%%%%%%%%%%

The paper is organised as follows. In Sect.~2 we present the observational material and describe the data reduction, including the artificial stars experiments. In Sect.~3 we will derive the Luminosity Function at different radii and we will show that it is the same everywhere. Sect.~4 is devoted to obtain an estimate of the stellar mass-to-light ratio by means of comparison with different theoretical models. The main results of the analysis are summarised in Sect.~5.

\section{Observations and Data Reductions}
\label{obs}

The data-set used for the present analysis consists of images obtained with the UVIS channel 
of the Wide Field Camera 3 (WFC3) aboard the HST. These images are part of a photometric calibration program  (Prop. ID: 11903; PI: J. Kalirai)
 so in principle they are not intended for LFs studies. Nevertheless thanks to the way in which they have 
 been planned and
 the extraordinary efficiency of the camera, 
 they result to be the deepest images available for NGC~2419. \\
 Given the aim of the present analysis we used only $F606W$ and $F814W$ images which allow one
 to reach fainter magnitudes than images with similar exposure times taken in different passbands and a straightforward comparison with theoretical models.
  The data-set used is thus composed by two images for each
 filter with the same exposure time $t_{exp}=400~s$ for $F606W$ and $t_{exp}=650~s$ for $F814W$.
 Images are perfectly aligned, with the cluster centre lying in chip $\#1$, at about $10\arcsec$ 
 from the gap which splits the camera in two twin chips.\\
 The photometric analysis has been performed on calibrated (i.e, bias and dark subtracted,
 and flat-field corrected) single exposure 
 images ({\it $\_$flt}) processed by the standard HST pipeline {\rm calwf3}.
 To these images we applied the most updated Pixel Area Map (PAM) corrections by using standard IRAF tasks.\\
 In order to obtain a reference frame cleaned of cosmic-rays and with an enhanced signal to
 noise ratio, we combined all the
 images in our data-set with the IRAF task {\em imcombine} applying a
 cosmic-ray rejection algorithm.\\
 For each filter and chip,
 a large number of relatively bright and isolated stars have been selected in order to 
 model properly the Point Spread Function (PSF) all over the Field of View. The PSF was modeled with a \citet{moffat} function, the parameters of which ($\sigma, \beta$) have been allowed to vary
linearly as a function of position.
 
 We used {\rm DAOPHOTII/ALLSTAR} \citep{s87} to search for sources at $2\sigma$ above 
 the background on the cleaned reference frame, thus obtaining a master list of bona fide stars.
Then, at the corresponding positions of these stars in the single images, a fit
was attempted with DAOPHOTII/ALLFRAME. 
For the stars recovered at the end of the procedure and for each filter the resulting magnitudes
were averaged and centroids positions combined by using DAOMATCH and DAOMASTER.\\

 The instrumental magnitudes thus obtained were transformed into the VEGAMAG photometric system using
 prescriptions and photometric zeropoints reported in the HST
 web-site.\footnote{http://www.stsci.edu/hst/wfc3/phot$\_$zp$\_$lbn}\\
 In order to assign absolute coordinates to each star of our final catalog, we corrected for the
 heavy geometric distortions which affects the WFC3, by applying the equations by \citet{bb09}.
 We used the stars in the data-set by D08 as secondary astrometric standards. We used {\rm CataXcorr}\footnote{CataXcorr is a code aimed at cross-correlating catalogues and finding astrometric solutions, developed by P. Montegriffo at INAF - Osservatorio Astronomico di Bologna, and successfully used by our group for the past 10 years.} to transform the WFC3 
 instrumental coordinates into Equatorial coordinate systems by using a few thousand stars 
 in common with D08.

%%%%%%%%%%%%%%%%%%%%%%%%%%%%%%%%%%%%%%%%%%%%%%%%%% FIG 
\begin{figure}
\includegraphics[width=80mm]{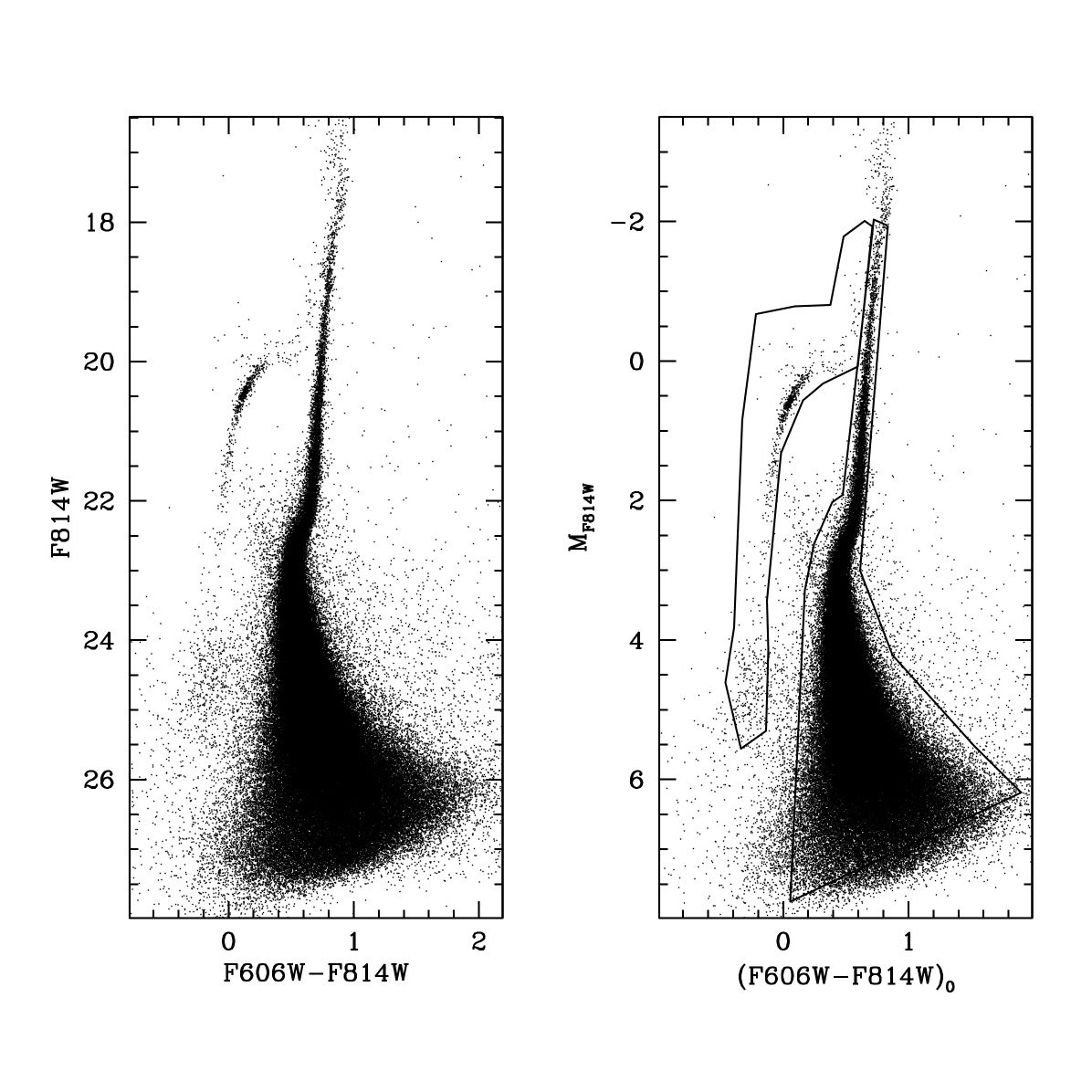}
 \caption{Colour Magnitude Diagrams of NGC~2419 from HST-WFC3 data, in the VEGAMAG system.
 Left panel: observed CMD. Right panel: CMD corrected for distance and reddening. The box adopted to select RGB+MS stars and HB stars are plotted as continuos lines.
 }
 \label{cmsel}
\end{figure}
%%%%%%%%%%%%%%%%%%%%%%%%%%%%%%%%%%%%%%%%%%%%%%%%%%%%%%%%%%%%%%%%%%%%%%%

In the left panel of Fig.~\ref{cmsel} we present the Color Magnitude Diagram (CMD) for the whole sample. The cluster Main Sequence dominates the diagram, with a Turn Off (TO) point around $F814W\sim 23.5$ and reaching the limiting magnitude of the photometry at $F814W\sim 27.5$. A sparse plume of BSS is clearly visible around $F606W-F814W\sim 0.0$ and $F814W\sim 22.5$. Stars on the steep Red Giant Branch, typical of metal-poor clusters, are reliably measured up to $F814W\sim 17.5$, where saturation sets in. The extended and bimodal HB reaches $F814W\sim 25.5$ \citep[see D08,][for details and discussion]{dic_mult}. In the right panel of Fig.~\ref{cmsel} magnitudes have been corrected for distance and reddening, and we show the selection boxes that we adopted to select likely cluster members for the RGB+MS and HB+AGB (Asymptotic Giant Branch). We use RGB+MS stars to derive the cluster LF in Sect.~3, while the HB+AGB sample is used in Sect.~4.1 to estimate the impact of these stars on the global mass-to-light ratio.

All over the paper we will consider the sample of stars having the DAOPHOT {\em sharpness} parameter in the range $-0.3<SHARP<0.3$; this selection criteria is found to clean the sample from sources whose shape is clearly different from the bulk of genuine stars without significantly affecting the overall sample size. Among the 156281 stars lying within the RGB+MS selection box, 139209 fulfil the  $-0.3<SHARP<0.3$ criterion. 

The contamination by foreground Galactic stars is expected to have little impact over such a small field of view as considered here
($162\arcsec\times162\arcsec\simeq0.002$~deg$^2$).
The predictions of the TRILEGAL Galactic model \citep{trilegal} and our analysis of the stellar population in the external region (beyond $R\simeq 600\arcsec$) of the 
D08 field indicates that less than ten Galactic stars are expected to contaminate our sample, all of them being brighter than $F814W\simeq 23.4$ ($M_{F814W}\simeq 3.5$). In particular, from the D08 dataset, we estimate that the fraction of Galactic stars in the sample selected for the analysis of the LF in the range $17.9\la F814W\la 23.4$ ($-2.0<M_{F814W}\le 3.5$) is less than 0.01 per cent. At fainter magnitudes unresolved background galaxies are, by far, the main source of contamination. Form the D08 dataset we estimate that the contamination by unresolved galaxies in the range $23.4\la F814W\la 25.4$ ($3.5<M_{F814W}\le 5.5$) is about 0.2 per cent. This may slightly under-estimate the actual contamination because it is not corrected for completeness, since we do not have such correction for the D08 sample; on the other hand it may overestimate the contamination
because many of these sources that appear star-like in the Subaru images used by D08 would be recognised as non-stellar in the WFC3 images. The limiting magnitude of the D08 sample (see Fig.~\ref{subaru}, below) prevented us to study the contamination fainter than $F814W\simeq 25.4$. From the publicly available catalogue of the HST Ultra Deep Field\footnote{\tt http://archive.stsci.edu/pub/hlsp/udf/acs-wfc/h\_udf\_wfc\_V1\_cat.txt} \citep{udf} we estimated that, for $F814W> 25.4$ to the limiting magnitude of our photometry,  the contamination by background galaxies is lower than a few per cent at any magnitude, i.e. smaller than the observational uncertainties in all the LFs we present in the following analysis.

\subsection{Artificial Stars Experiments}

A crucial step in determining an accurate LF is a proper estimation 
of photometric completeness. This aspect is particularly important in the case of NGC~2419, which appears to be 
a remarkably crowded cluster because of its distance. 
 
\begin{figure}
\includegraphics[width=80mm]{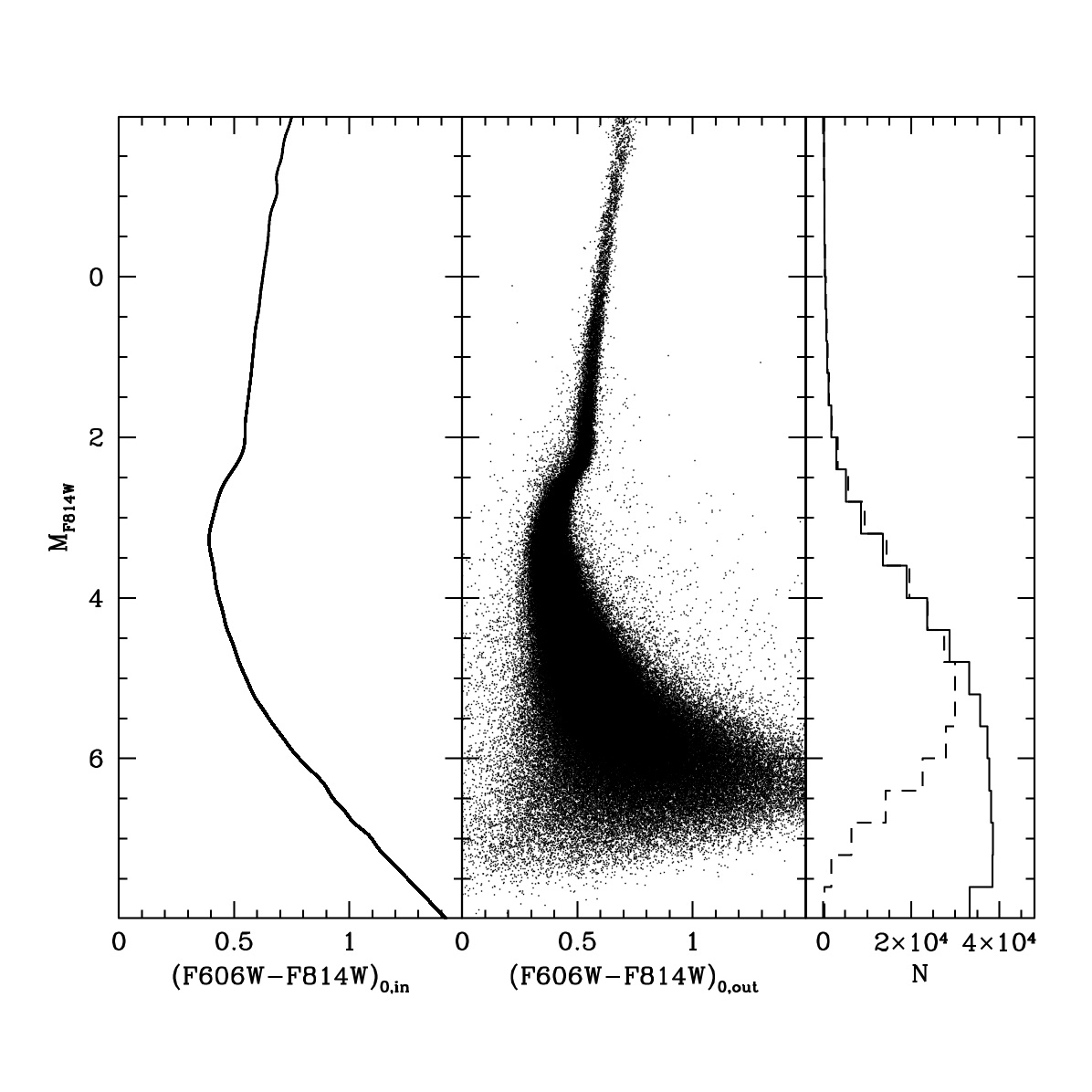}
 \caption{Colour Magnitude Diagrams and Luminosity Functions for artificial stars.
 Left panel: CMD from input magnitudes. Middle Panel: CMD for output magnitudes.
 Right panel: Luminosity Function for the {\em input} (continuous histogram) and {\em output} (dashed histogram) samples.
 }
 \label{cmsim}
\end{figure}

Artificial stars experiments were performed 
following the method described by Bellazzini et al. (2002) and \citet{giacomo}.
We generated a catalog of simulated stars
with a $F814W_{in}$ magnitude extracted from a LF modelled to reproduce the observed LF in that band and extrapolated beyond the limiting magnitude \citep[see][]{n288}. 
Then to each star extracted from the LF we assigned a $F606W_{in}$ magnitude by means of
an interpolation 
along the mean ridge line of the cluster (left panel of Figure~\ref{cmsim}).

Artificial stars were added to real images (including the reference frame) by using the {\rm DAOPHOTII/ADDSTAR} software.
In order to avoid ``artificial crowding", stars were placed into the images following a regular grid
defined by 15 pixels cells (6-7 times the usual FWHM of stars for these images) in which only one artificial 
star for each run was allowed to lie.

The photometric reduction process used for the artificial stars experiment is exactly the same as described in
Section~\ref{obs}. Those stars recovered after the photometric analysis have also a $F606W_{out}$ and $F814W_{out}$. 
 More than 200000 stars have been simulated for each chip. The photometric completeness
$C_f$ is defined as the ratio between the number of stars recovered at the end of the procedure and  
the total number of stars actually simulated.   

The {\em input} and {\em output} CMDs and LFs of the artificial stars sample are compared in Fig.~\ref{cmsim}. 
Note the close similarity of the synthetic CMD with real one plotted in Fig.~\ref{cmsel}.
Throughout the paper we always adopt {\em the same selection criteria for observed and artificial stars.}

\begin{figure}
\includegraphics[width=80mm]{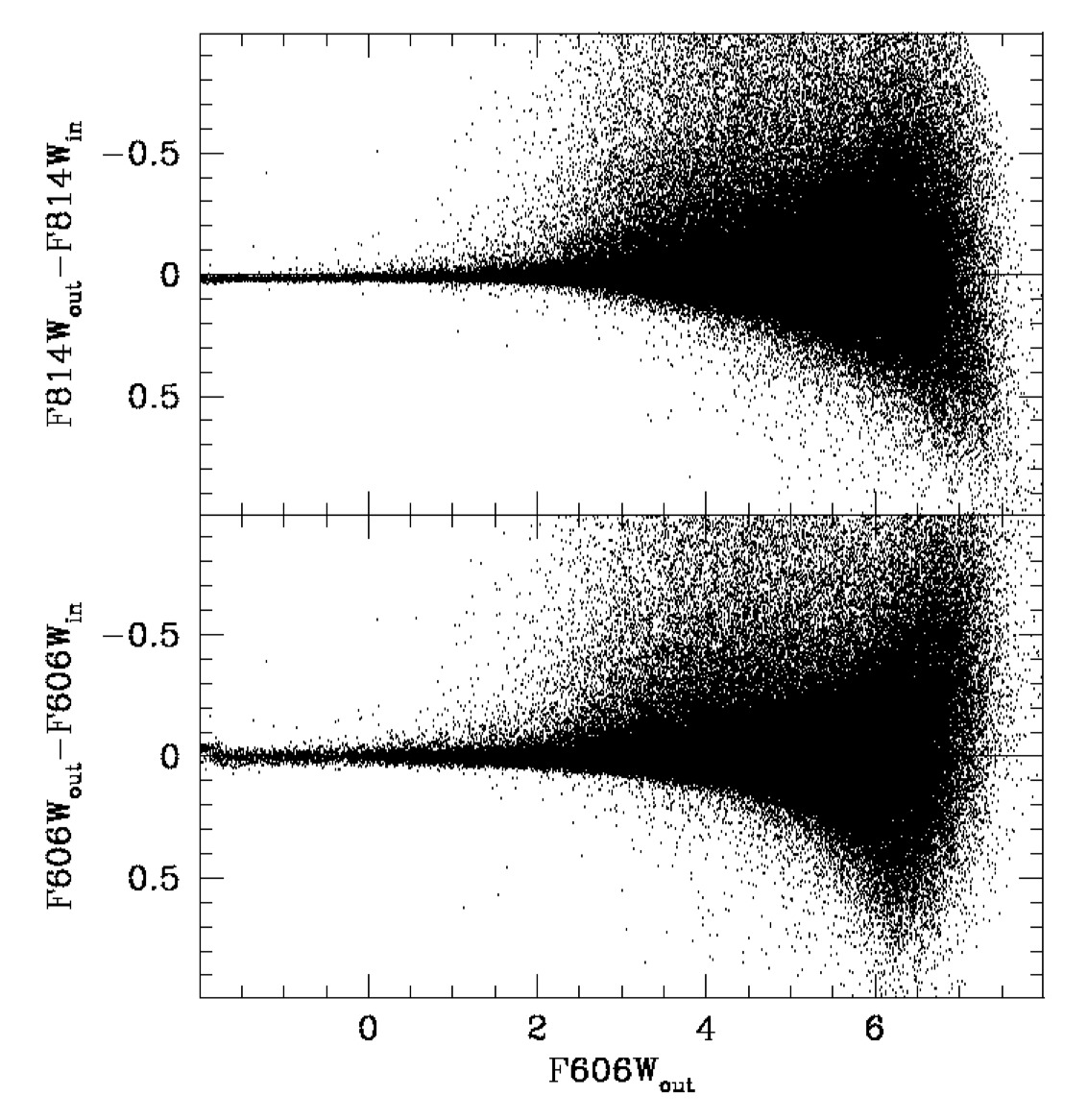}
 \caption{Distribution of the input-output magnitude differences, as a function of output $M_{F814W}$ magnitude.
 }
 \label{errcf}
\end{figure}

In Figure~\ref{errcf} we compare the residuals between {\em input} and {\em output} magnitudes for both filters.
As expected, the distribution is not symmetrical: a significant number of stars have been recovered with
 an {\em output} magnitude brighter than the assigned {\em input} of a quantity larger of their typical 
 photometric error. This effect is due to those stars that blended with real stars with similar
 (or larger) luminosity. This effect has been taken into account properly when building the LF (see Section~\ref{lf}). 
 
%%%%%%%%%%%%%%%%%%%%%%%%%%%%%%%%%%%%%%%%%%%%%%%%%% FIG 
%%%%%%%%%%%%%%%%%%%%%%%%%%%%%%%%%%%%%%%%%%%%%%%%%%%%%%%%%%%%%%%%%%%%%%%

%%%%%%%%%%%%%%%%%%%%%%%%%%%%%%%%%%%%%%%%%%%%%%%%%% FIG 
%%%%%%%%%%%%%%%%%%%%%%%%%%%%%%%%%%%%%%%%%%%%%%%%%%%%%%%%%%%%%%%%%%%%%%%

\subsection{Completeness Functions}
\label{lf}

In the following we will consider the sample of stars enclosed in the MS+RGB selection box displayed in Fig.~\ref{cmsel} and the analysis will be performed using absolute magnitudes ($M_{F814W}$).

\begin{figure}
\includegraphics[width=80mm]{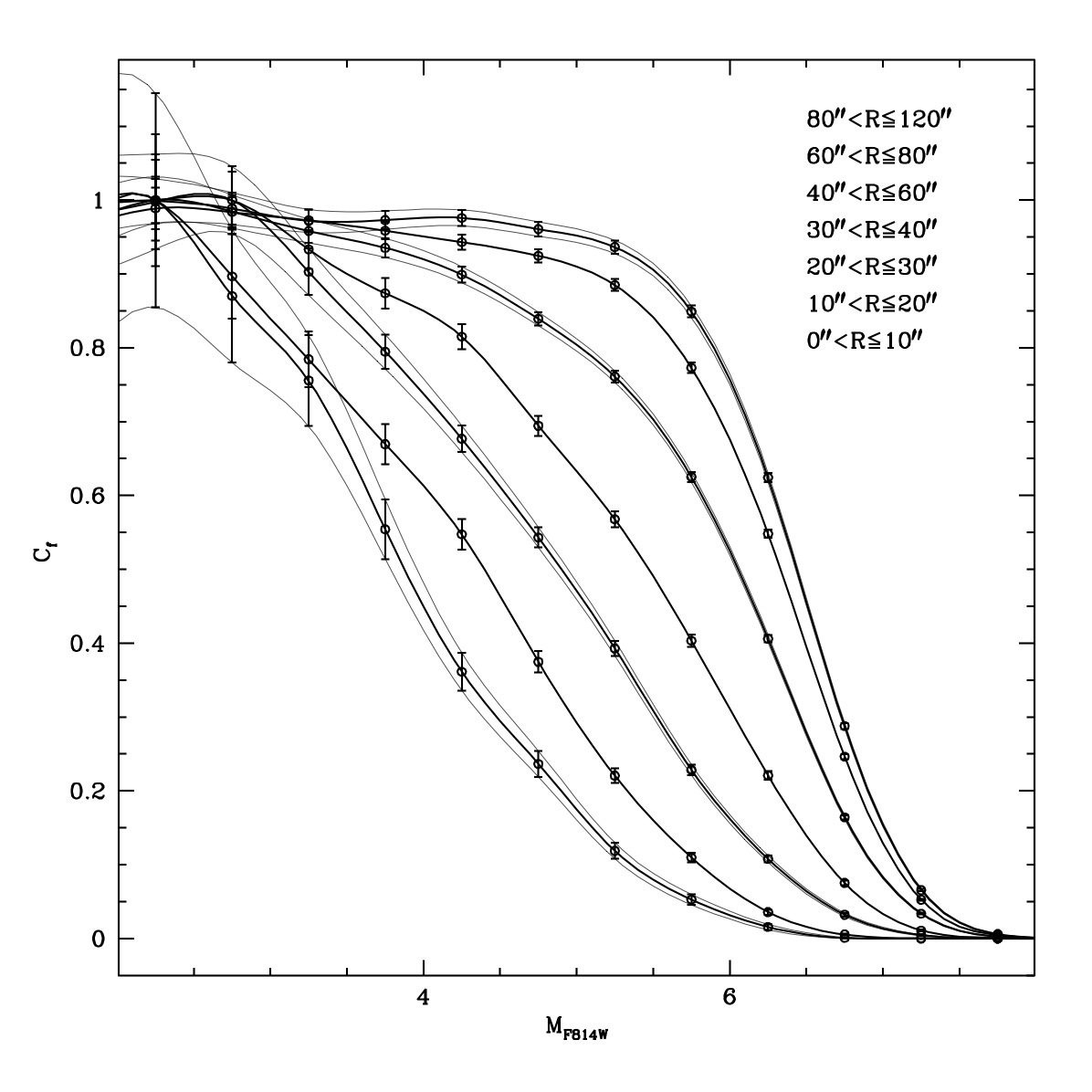}
 \caption{Completeness as a function of magnitude for several radial ranges (the leftmost curve is for $0\arcsec<R\le 10\arcsec$, the rightmost is for $80\arcsec<R\le 120\arcsec$. The lines connecting the points are spline interpolations through the data. The lines through the edges of the error-bars (reported for the completeness functions of four radial ranges, as examples) show how the uncertainties can be reliably interpolated with splines. 
 }
 \label{VEDICF}
\end{figure}

In Fig.~\ref{VEDICF} we plot the completeness factors as a function of magnitude for several ranges of distance from the cluster centre (R). The crucial conclusion that can be drawn from this plot is that  at any given magnitude fainter than $M_{F814W}= 2.0$, the degree of completeness is subject to significant variations on scales as small as $\sim 10\arcsec$. Conversely, for $M_{F814W}\le 2.0$, the completeness is 100\% at any radius; this is true also for $M_{F814W}\le 3.0$, at least for $R>20.0\arcsec$. Usually, in nearby GCs, the completeness is nearly constant over relatively large radial intervals, so that a unique correction can be applied to the LF from stars in that range \citep[see, e.g.][]{n288}. This is clearly not the case for the remote cluster considered in this study. To account for the fast variation of $C_f(m)$ with radius and for the effects of bin migration along the LF due to blending convolved with photometric error (see Fig.~\ref{errcf}), we adopted the following  correction procedure acting on a star-by-star basis, inspired by \citet{bai} and \citet{n288}:

\begin{enumerate}

\item For each star with $M_{F814W}> 2.0$, and having $M_{F814W}=m_0$ and $R=r_0$, we derive the Completeness as a function of magnitude, with the associated uncertainties, from the sample of artificial stars lying within $r_0-5.0\arcsec<R<r_0+5.0\arcsec$, with the same binning as in Fig.~\ref{VEDICF}.

\item We interpolate on the derived {\em local} $C_f(m)$\footnote{That would be more properly denoted as $C_f(m,r)$, to emphasise its local nature. However we prefer to drop the $r$ argument to avoid a too heavy notation.} with splines to obtain the completeness and the associated uncertainty at $m_0$, $C_f(m_0)$ and  $ errC_f(m_0)$. The construction of the whole $C_f(m)$ allows a more reliable (less noisy) estimate of $C_f(m_0)$ than extracting only the stars in the magnitude bin of the considered star, since the overall shape of the $C_f(m)$ contributes to the derivation of the actual $C_f(m_0)$ value.

\item We extract at random a Gaussian deviate $G_{dev}$ and we assign to the star the completeness factor $C_{f,\star}=C_f(m_0)+G_{dev}\times errC_f(m_0)$, with the additional constraints that if the value of $C_{f,\star}$ exceeds 1.0 its is set equal to 1.0, and if it is lower than 0.0 it is set to 0.0, so that $0.0\le C_{f,\star}\le 1.0$. This step takes into account the effect of the uncertainty in the determination of $C_f(m_0)$.

\item The subsample of stars having $m_0-0.5<m_{out}<m_0+0.5$ is extracted from the sample of artificial stars lying within $r_0-5.0\arcsec<R<r_0+5.0\arcsec$. We compute $\Delta m=m_{out}-m_{in}$ for these stars and we chose one of them at random. Its $\Delta m$ is then assigned to the considered real star ($\Delta m_{\star}$). The magnitude of the real star is then corrected by the effect of bin-migration (blending + photometric error) by subtracting $\Delta m_{\star}$ to its observed magnitude, i.e. $m_{\star}=m_0-\Delta m_{\star}$. While the correction it is clearly unjustified on an individual star basis, on a large sample as the one considered here, it should provide a sound {\em statistical} correction.

Once the above four steps are performed for all the stars in the sample, each observed star has a migration-corrected magnitude $m_{\star}$ and an individual completeness value $C_{f,\star}$. Both of them have been derived from stars lying in the same radial and magnitude range of the considered star, so that the variations of these effects as a function of $m$ and R are properly accounted for.
The only limitation is due to the adopted size of the radius and magnitude bins, that should be sufficiently large to ensure robust determination of $m_{\star}$ and $C_{f,\star}$.

\item Now the production of the corrected LF for any radial range is straightforward. One has simply to divide the stars lying in that range in magnitude bins, according to their $m_{\star}$, and then, instead of counting the stars falling within each bin $i$, the inverse of their completeness factors must be summed $N_i=\sum{\frac{1}{C_{f,\star}}}$ \citep[see][]{bai}. For each bin we also record the average $C_f$ and we retain for the final LF only bins having $\langle C_f\rangle\ga 0.5$.

\item To minimise the shot noise introduced by the random extractions included in the procedure we produced ten realisations of the catalog with associated $m_{\star}$ and $C_{f,\star}$. Each LF presented below is obtained by averaging each $N_i$ over the ten realisations. The associated standard deviation is added in quadrature to the Poisson noise of each bin to obtain a final uncertainty on $\langle N_i\rangle$.

\end{enumerate}

%In the next section we present and discuss the derived LFs.

%%%%%%%%%%%%%%%%%%%%%%%%%%%%%%%%%%%%%%%%%%%%%%%%%% FIG 
%%%%%%%%%%%%%%%%%%%%%%%%%%%%%%%%%%%%%%%%%%%%%%%%%%%%%%%%%%%%%%%%%%%%%%%

\section{The Luminosity Function}

\begin{figure}
\includegraphics[width=80mm]{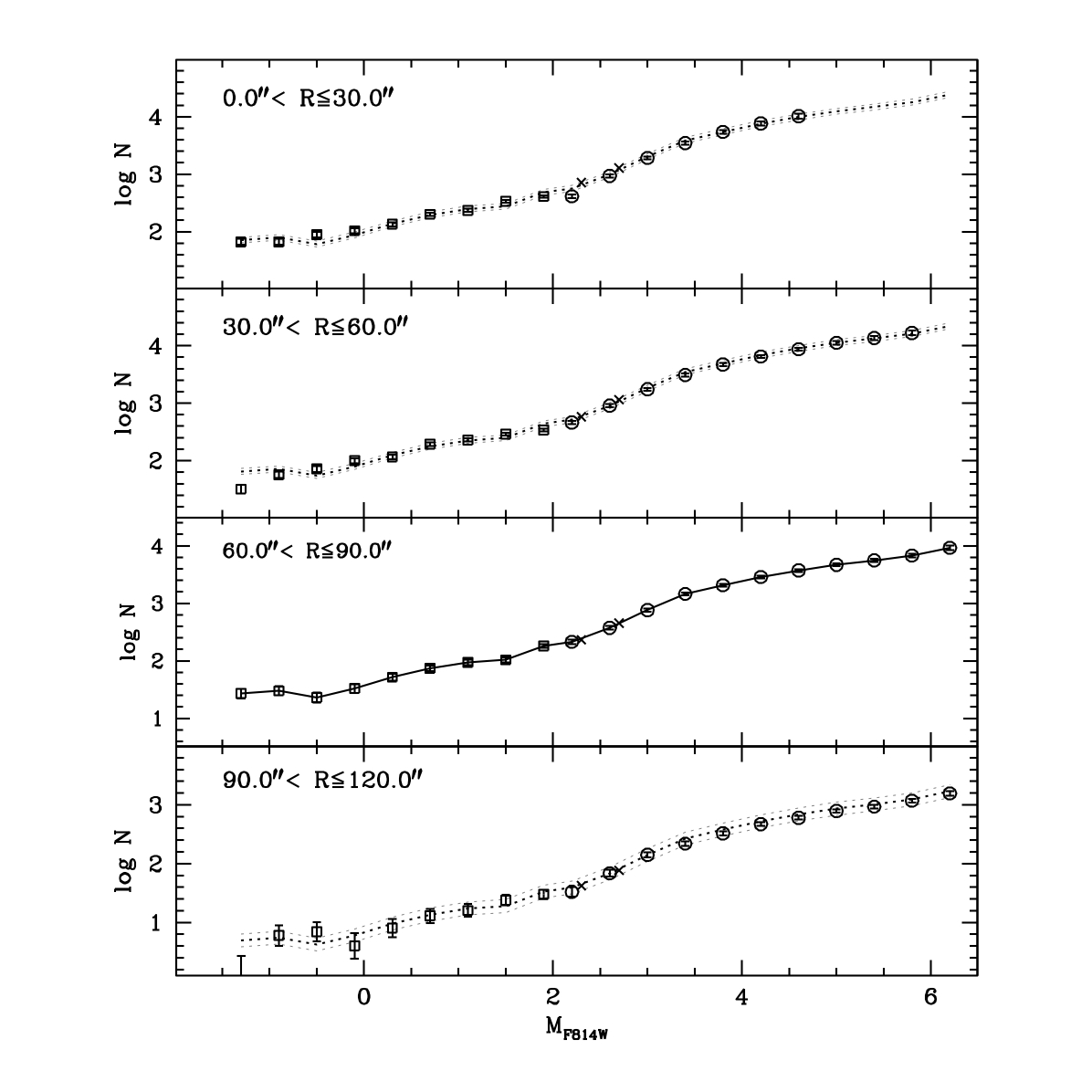}
 \caption{Luminosity Function of NGC~2419 for four radial ranges (note that $R=60\arcsec$ corresponds approximately to the half-light radius of the cluster).
The faint part of the LF  (open circles) have been obtained applying the full Completeness Correction, the bright part (open squares) refers to the magnitude range where $C_f=1.0$ at any radius. The two faintest points of the bright LF ($\times$ symbols) overlap with the faint LF, showing that the two branches of the LF are fully compatible, i.e. there is no mismatch in the normalisation. The continuous line connecting the data points of the $60\arcsec<R\le90\arcsec$ is 
compared (dotted lines) to the LFs in all the other panels, after normalisation to the four points in the range $0.5<M_{F814W}<2.0$. The thin dotted lines enclose the $\pm 2\sigma$ range of the uncertainty on the normalisation factor.
 }
 \label{complf}
\end{figure}

In Fig.~\ref{complf} we show the corrected LFs in four radial ranges sampled by our WFC3 data. These approximately correspond to 0.0-0.5$r_h$, 0.5-1.0$r_h$, 1.0-1.5$r_h$, and 1.5-2.0$r_h$, from top to bottom, respectively. The {\em faint branch} of the LFs ($M_{F814W}> 2.0$), plotted as open circles, have been derived as described above, while for the {\em bright branch} ($M_{F814W}< 3.0$, open squares) we produced the LFs as ordinary histograms since, as mentioned above, in the $M_{F814W}<2.0$ range $C_f=1.0$ at all radii. We extended the bright branch down to $M_{F814W}\la 3.0$ ($C_f=1.0$ anywhere except for the innermost regions, where it is $\ga 90$\%) to verify that the {\em faint} and {\em bright} branches are in agreement in the region of overlap.
This is indeed the case {\em without any adjustment of the normalisation}, thus demonstrating that the adopted correction procedure gives self-consistent results.

The LFs reaches different magnitude limits on the faint end. Due to increasing crowding toward the centre, the $\langle C_f\rangle=0.5$ threshold occurs at brighter magnitudes in  the inner regions. The LF in the range $60\arcsec<R\le 90\arcsec$ reaches the same limit as the outermost LF ($M_{F814W}=6.2$ at the bin centre, implying the inclusion of stars down to
$M_{F814W}=6.4$, that is the faintest limit of our LFs) but has been obtained from a much larger sample. For this reason we take it as a {\em reference} for comparison with LFs in other radial ranges and we list it in Tab.~\ref{tablf}. 
The line connecting the points of the reference LF is plotted as a dotted line in the panels showing the LF in the other radial ranges in Fig.~\ref{complf}, after normalisation to the four points in the range $0.5<M_{F814W}<2.0$. This approach to normalisation is applied in all cases of comparison between empirical or theoretical LFs in the following analysis.

Our LF samples the cluster population from $\simeq 2$~mag above the HB to 2-3~mag below the Main Sequence TO, depending on the considered radial range. Adopting an isochrone from the Dartmouth set \citep{dotter}, with [Fe/H]=-2.1, [$\alpha$/Fe]=+0.4, Y=0.2455 and age = 12~Gyr (that provides a good fit to the observed CMD, see Fig.~\ref{cm}, below), the magnitude range covered by the {\em reference} LF corresponds to the mass range between 0.81~$M_{\sun}$ and 0.53~$M_{\sun}$.
The key result of Fig.~\ref{complf} is that, in the range of magnitudes/masses covered by our LFs, {\em there is no difference in the LFs obtained in different radial ranges}, over $0.0r_h<R<2.0r_h$, or, equivalently,  $0.0r_c<R<6.0r_c$. The shape of various LFs is clearly the same and any subtle difference lies well within the combined uncertainty of the LFs and of the adopted normalisation.

%%%%%%%%%%%%%%%%%%%%%%%%%%%%%%%%%%%%%%%%%%%%%%%%TABELLA LF60-90
\begin{table}
\label{tablf}
 \centering
 \begin{minipage}{70mm}
  \caption{Luminosity Function of NGC~2419 in the range $60\arcsec<R\le 90\arcsec$.}
  \begin{tabular}{@{}cccc@{}}
  \hline
     $M_{F814W}$  & log N\footnote{Computed on 0.4 mag bins.} & err(log N)  
     & branch\footnote{1 = bright branch; 2 = faint branch.}\\
  \hline

        -1.3   &    1.431  &   0.083 & 1 \\
        -0.9   &    1.477  &   0.079 & 1 \\
        -0.5   &    1.362  &   0.090 & 1 \\
        -0.1   &    1.519  &   0.076 & 1 \\
         0.3   &    1.716  &   0.060 & 1 \\
         0.7   &    1.869  &   0.050 & 1 \\
         1.1   &    1.973  &   0.045 & 1 \\
         1.5   &    2.017  &   0.043 & 1 \\
         1.9   &    2.260  &   0.032 & 1 \\
         2.2   &    2.331  &   0.039 & 2 \\
         2.6   &    2.574  &   0.032 & 2 \\
         3.0   &    2.882  &   0.028 & 2 \\
         3.4   &    3.163  &   0.026 & 2 \\
         3.8   &    3.314  &   0.025 & 2 \\
         4.2   &    3.458  &   0.025 & 2 \\
         4.6   &    3.571  &   0.025 & 2 \\
         5.0   &    3.670  &   0.025 & 2 \\
         5.4   &    3.746  &   0.026 & 2 \\
         5.8   &    3.830  &   0.029 & 2 \\
         6.2   &    3.966  &   0.038 & 2 \\
\end{tabular}
\end{minipage}
\end{table}
%%%%%%%%%%%%%%%%%%%%%%%%%%%%%%%%%%%%%%%%%%%%%%%FINE TABELLA LF60-90
	 
%%%%%%%%%%%%%%%%%%%%%%%%%%%%%%%%%%%%%%%%%%%%%%%%%% FIG 
%%%%%%%%%%%%%%%%%%%%%%%%%%%%%%%%%%%%%%%%%%%%%%%%%%%%%%%%%%%%%%%%%%%%%%%
			    
\begin{figure}
\includegraphics[width=80mm]{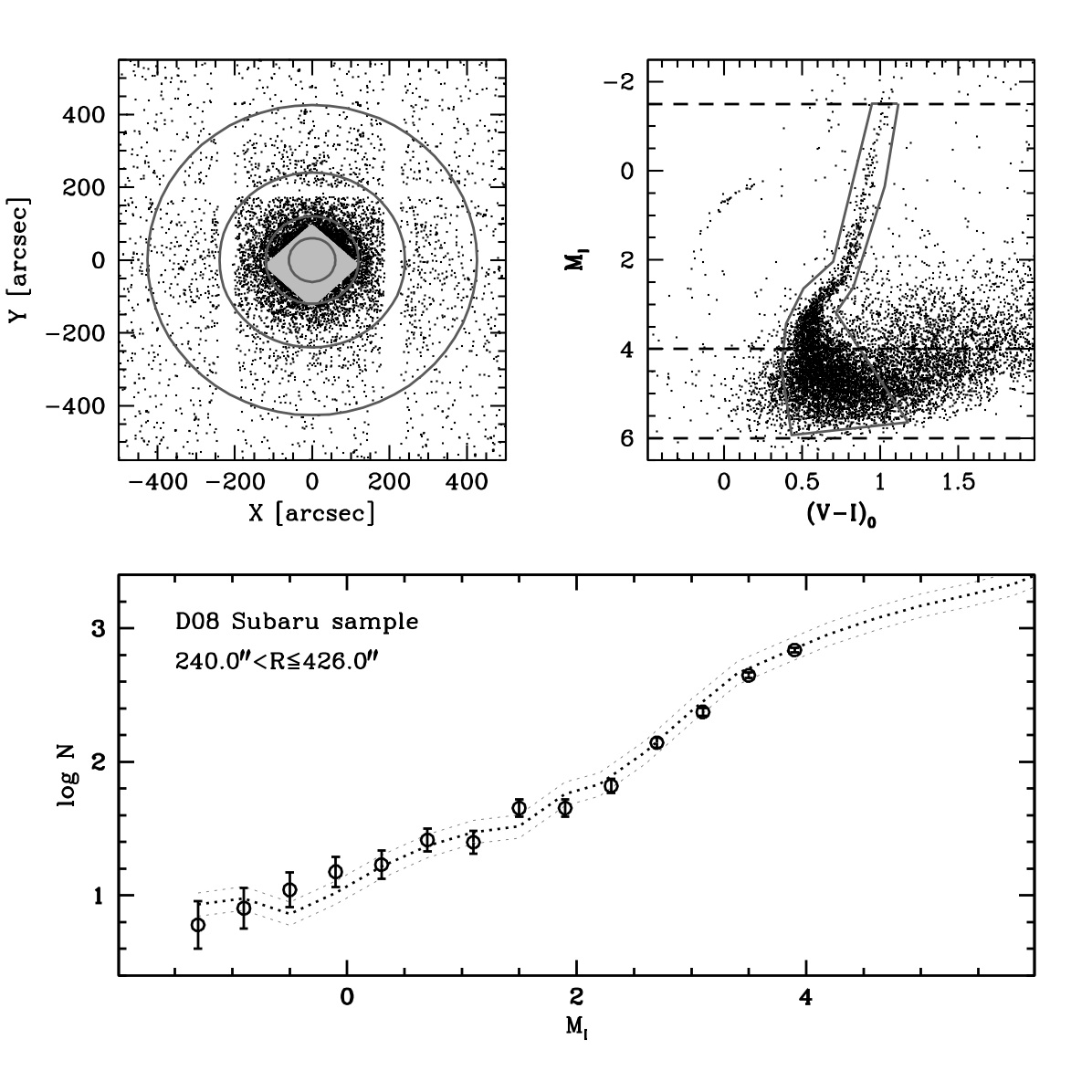}
 \caption{
Upper left panel: map of the combined ACS + SUBARU-Suprimecam sample by D08 (black dots; only stars with $M_I\le 4.0$ have been plotted) with the WFC3 sample superposed (light gray area). The circles are centred on the centre of the cluster and have radius=$60\arcsec$, $120\arcsec$, 
$240\arcsec$, and $426\arcsec$. 
Upper right panel: CMD from the same sample with the box adopted to select cluster stars for the LF. The long dashed lines mark the limiting magnitude and $M_I=4.0$, two magnitudes above that level. Lower panel: LF for cluster stars with $M_I<4.0$ in the radial range corresponding to $4.3r_h\la R\la 7.6r_h$. The dotted line is the LF from the WFC3 sample in the range 
 $60\arcsec< R\le 90\arcsec$ (see Fig.~\ref{complf}), after normalisation to the four points in the range $0.5<M_{F814W}<2.0$. The thin dotted lines enclose the $\pm 2\sigma$ range of the uncertainty on the normalisation factor. 
 }
 \label{subaru}
\end{figure}

The comparison presented in Fig.~\ref{complf} suffers from two main limitations. The first is that a large range of stellar masses is beyond the reach of our LFs, i.e. from 0.5~$M_{\sun}$ to the hydrogen burning limit at $\simeq 0.1~M_{\sun}$. Given the large distance of NGC~2419 this would be very difficult to overcome. A simple experiment with the WFC3 Exposure Time Calculator (ETC) shows that to detect stars two magnitudes fainter than our limit with S/N=3, more than 70~h of exposure time per filter are required. Moreover, pushing the limit of the LF two mag fainter than ours would still imply a mass limit of $0.3~M_{\sun}$, still significantly above the H burning limit. Hence, the extension of the LF/MF to a significantly larger mass range seems impractical, at the present time. The second limitation arises from the radial range covered by the WFC3 data, since the clusters extends at least out to $r_t=426\arcsec \simeq 7.6~r_h$ \citep{mic}\footnote{The best-fit model by \citet{iba11a} has $r_t=781\arcsec$, however, beyond $R=410\arcsec$ the Surface Brightness profile fall below $\mu_V\simeq 29.0$ mag/arcsec$^2$, implying a surface density of stars that is probably too low to derive statistically sound LFs with HST instruments.}. This can be partially overcome by using the deep wide-field
ground-based photometry obtained by D08 from SuprimeCam@Subaru data, taking advantage of the equivalence between F814W and I photometry that we discussed in Sect.~1. In the upper panels of Fig.~\ref{subaru} we display the adopted selections for this sample. Since in the regime of moderate crowding the completeness typically reaches its flat branch about 2 magnitudes above the limiting magnitude of the sample, we considered only stars lying above this limit ($M_I<4.0$) and we derived the LF without any completeness correction. We focussed on the radial range $240\arcsec<R\le 426\arcsec$ corresponding to $4.3r_h\la R\la 7.6r_h$, i.e. sufficiently external to avoid the most crowded region and sufficiently large to obtain a reliable LF of the cluster outskirts.

%%%%%%%%%%%%%%%%%%%%%%%%%%%%%%%%%%%%%%%%%%%%%%%%%% FIG 
%%%%%%%%%%%%%%%%%%%%%%%%%%%%%%%%%%%%%%%%%%%%%%%%%%%%%%%%%%%%%%%%%%%%%%%

In the lower panel of Fig.~\ref{subaru} we compare the LF from the Subaru data to our {\em reference} LF (dotted line), with the usual normalisation. Also in this case the LFs in these two widely different radial ranges clearly display the same shape. We verified that contamination from foreground sources and/or background unresolved galaxies does not significantly affect this comparison (see Sect.~2).
Hence, our analysis of the radial behaviour of the LF fully supports the conclusion reached by D08 using the BSS as tracer: the stars of NGC~2419 show no sign of mass-segregation all over the whole body of the cluster \citep[see][for discussion]{baum09,ema}. This evidence strongly supports the hypothesis of constant M/L with radius that has been adopted by \citet{baum09} and \citet{iba11a,iba11b} in their analysis of the cluster dynamics. 

However, there is a caveat that must be mentioned. Our conclusion is limited to light-emitting stars:
dark remnants much more massive than the average cluster star (like black holes - BH -  and neutron stars - NS) may be centrally segregated \citep{spitzer}, i.e. they may have a radial distribution different from the stars. The possibility that this would have a significant effect on the overall mass distribution does not seems likely but it cannot be excluded. Given the present day status of the cluster and using the Equation 34 by \citet{mouri}, we find that the mass-segregation timescale ($\tau_{ms}$) in NGC~2419 is as large as 11~Gyr for $m=1.4~M_{\sun}$ NS, and it becomes significantly smaller than the cluster age only for BHs with $m\ga 10~M_{\sun}$ 
($\tau_{ms}=1.6$~Gyr). Such high mass remnants should be quite rare and are not expected to provide large contribution to the overall mass budget. In Sect.~\ref{subml}
and Sect.~\ref{summary}, below we will discuss in more detail the contribution of dark remnants (including white dwarfs) to the total cluster mass.

\section{Comparison with theoretical models}
\label{teo}

To constrain the global mass-to-light ratio we adopt the following approach:

\begin{itemize}

\item We fit the observed LFs with theoretical counterparts from evolutionary models with appropriate physical parameters. There are now dedicated web tools that allow one to produce LFs of the desired form \citep[e.g. choosing the exponent $x$ of the power law IMF $dN\propto m^x dm$; where $x=-2.35$ corresponds to the popular IMF of][]{salp} from theoretical isochrones of given chemical composition and age, so it is easy to assemble a grid of theoretical LFs to compare with observations. 

\item The best fit theoretical LF can be numerically integrated to obtain the total mass and the total luminosity (in any given passband), thus obtaining the stellar M/L independent of the normalisation. The effect of dark remnants can be taken into account with suitable assumptions \citep[see][]{krui_remn}.

\end{itemize}

%%%%%%%%%%%%%%%%%%%%%%%%%%%%%%%%%%%%%%%%%%%%%%%%%% FIG 
\begin{figure}
\includegraphics[width=80mm]{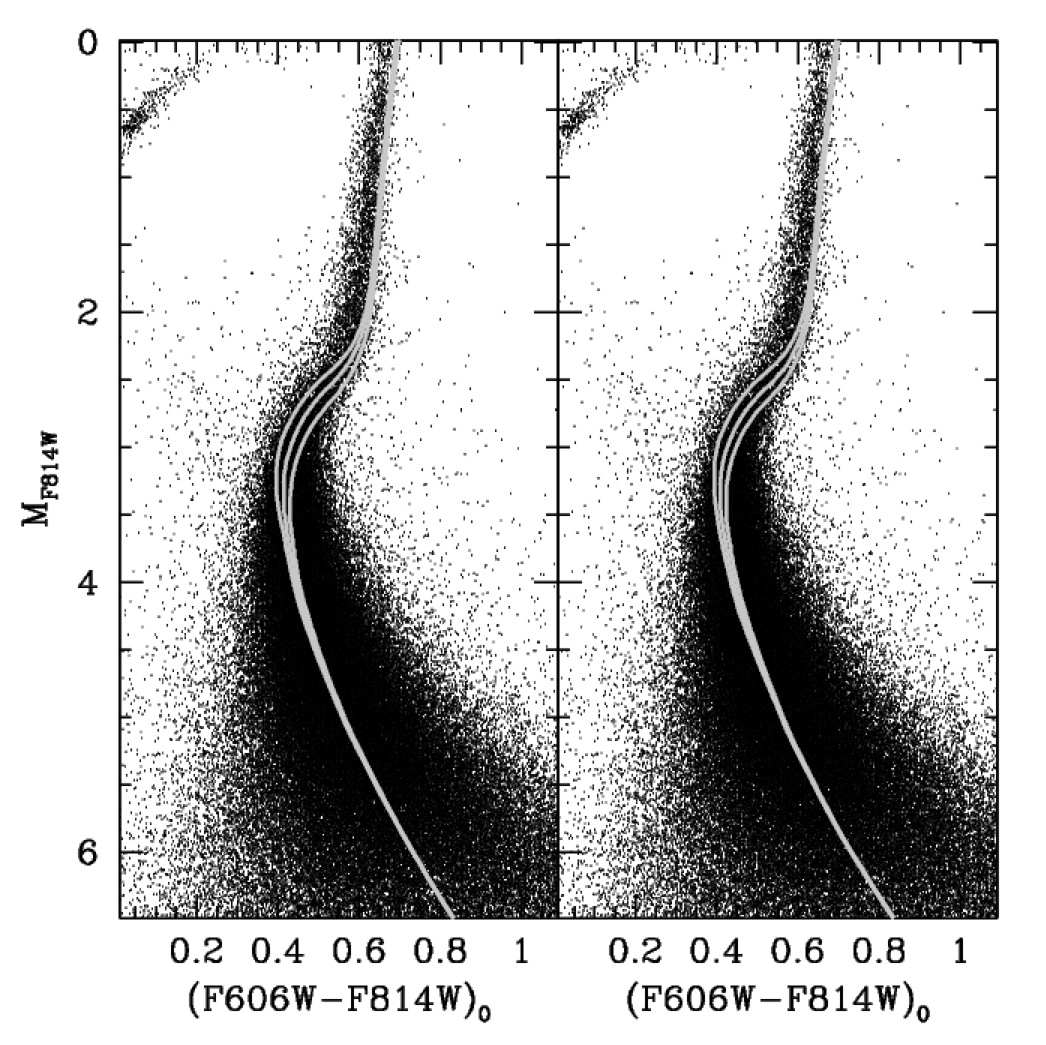}
 \caption{Comparison of the observed CMD with theoretical isochrones from the set by \citet{dotter} with [Fe/H]=-2.1. Left panel: isochrones with [$\alpha$/Fe]=+0.2, and age=12.0, 13.0, and 14.0 Gyr, from left to right. Right panel: isochrones with [$\alpha$/Fe]=+0.4, and age=11.0, 12.0, and 13.0 Gyr, from left to right. 
 }
 \label{cm}
\end{figure}
%%%%%%%%%%%%%%%%%%%%%%%%%%%%%%%%%%%%%%%%%%%%%%%%%%%%%%%%%%%%%%%%%%%%%%%

We adopt the Dartmouth theoretical models \citep{dotter}\footnote{\tt http://stellar.dartmouth.edu/$\sim$models/} as our reference set. There are several 
reasons for this choice: (a) theoretical LFs and isochrones are provided down to the H-burning limit, in the WFC3 system, (b) the desired LFs can be produced with a web tool\footnote{\tt http://stellar.dartmouth.edu/$\sim$models/isolf.html}, and (c) this is the same set used by \citet{paust09,paust10} in their exhaustive study of GCs mass functions, hence we have an homogeneous study to compare with. In the present context it is relevant to recall that \citet{paust09} showed that different sets of theoretical models provide MLRs for metal-poor stars in good agreement with each other, hence this is not expected to be a significant source of uncertainty in the determination of M/L ratios. 

In Fig.~\ref{cm} we compare theoretical isochrones from the Dartmouth set with the observed CMD of NGC~2419. All isochrones have $[Fe/H]=-2.1$, to match the spectroscopic estimate by \citet{judyhr}. Those in the left panel have $[\alpha/Fe]=+0.2$, while those on the right have $[\alpha/Fe]=+0.4$. The latter value is equal to the measured average reported in Table~\ref{tabpar}, however, since the average abundance of each individual $\alpha$ element can be quite different from this \citep[even if always $>0.0$][]{judyhr} it seems worth to explore the effect of different assumptions on $[\alpha/Fe]$. It is quite clear that satisfactory fits of the MS and TO regions of the CMD can be obtained with both sets of isochrones. The well known age-metallicity degeneracy manifests in the fact that a slightly younger isochrone is required to fit the CMD with the set with the larger overall metallicity (i.e. the one with $[\alpha/Fe]=+0.4$), 12 Gyr instead of 13 Gyr. We have verified that the choice between the two sets does not have any significant impact on the LF fit and the consequent M/L estimate. In the range of age and global metallicity $[M/H]$
\footnote{$[M/H]=[Fe/H]+{\rm log}(0.638\times 10^{[\alpha/Fe]}+0.362)$, where $[M/H]={\rm log}Z-{\rm log}Z_{\sun}$, see \citet{scs93,f99} and references therein. With the assumptions listed in Table~\ref{tabpar}, we obtain $[M/H]=-1.8$ for NGC~2419, corresponding to $Z=0.0003$.} that is relevant for the present case the effects of these parameters on the LF shape are quite weak, the dominant parameter is the exponent of the MF power law. In the following we will adopt $[Fe/H]=-2.1$, $[\alpha/Fe]=+0.4$ and age=12.0 Gyr as the {\em standard} set of values for the Dartmouth LFs to be compared with observations. 
When using other sets of theoretical models, the BASTI set \citep{basti}\footnote{\tt http://193.204.1.62/index.html} or the Padua set \citep{marigo}\footnote{\tt http://stev.oapd.inaf.it/cgi-bin/cmd}, we keep the global metallicity fixed and we slightly adjust the age to fit the observed LFs. We recall that the normalisation between observed and theoretical LFs is always performed on the four points in the range $0.5<M_{F814W}<2.0$. 

%%%%%%%%%%%%%%%%%%%%%%%%%%%%%%%%%%%%%%%%%%%%%%%%%% FIG 
\begin{figure}
\includegraphics[width=80mm]{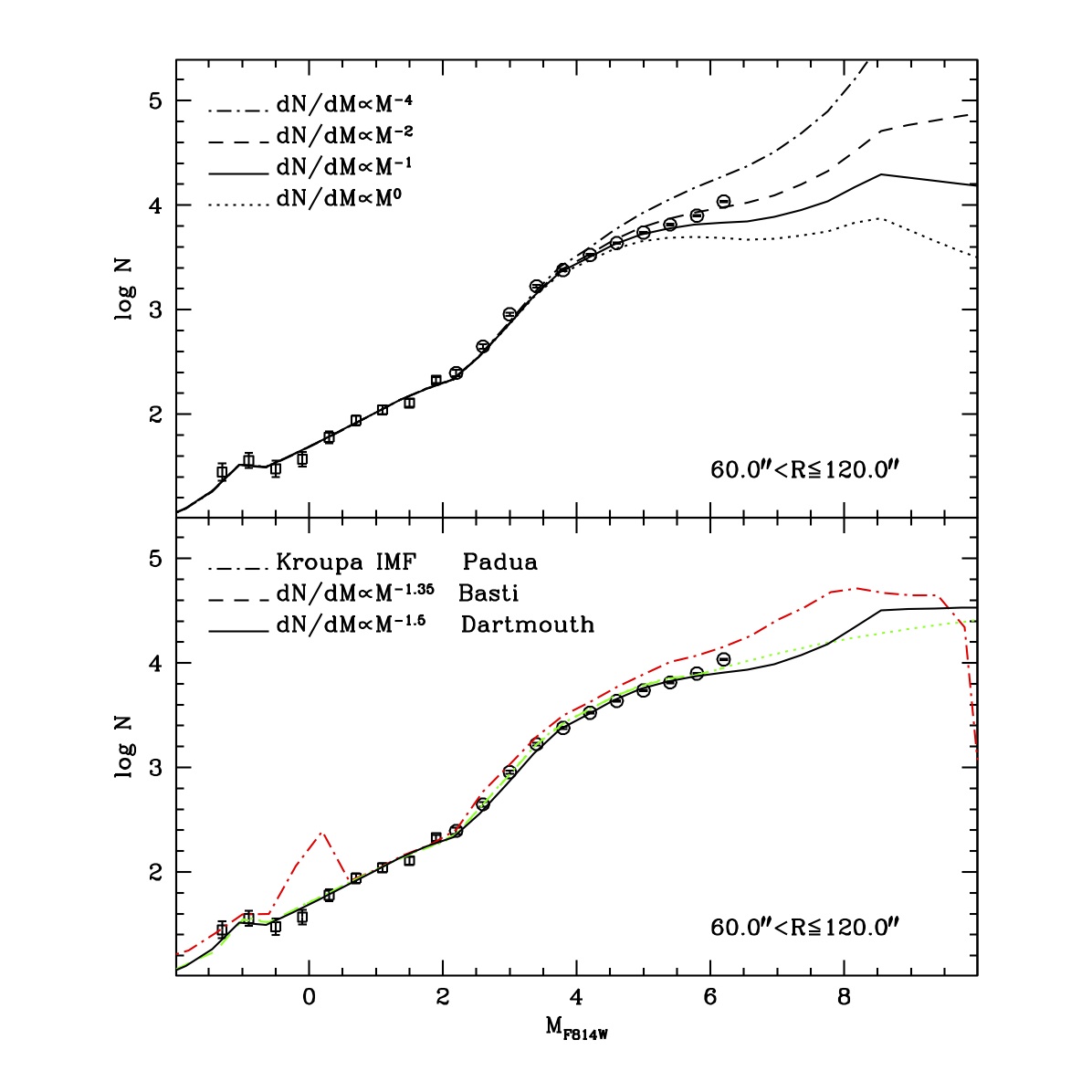}
 \caption{Observed LF compared with theoretical models, after normalisation to the four points in the range $0.5<M_{F814W}<2.0$. Upper panel: comparison with Dartmouth models; different line styles are adopted for IMFs with different exponents, as indicated in the upper left corner of the panel. All the LF models have [Fe/H]-2.1, [$\alpha$/Fe]=+0.4, age=12.0 Gyr. Lower panel: comparison with a Basti model with [Fe/H]=-1.8, [$\alpha$/Fe]=0.0, age=13.0 Gyr and IMF exponent =-1.35 (long dashed line) extrapolated to the H-burning limit with a spline (dotted line, see text for details), and with a Padua model with [Fe/H]=-1.8, [$\alpha$/Fe]=0.0, age=13.5 Gyr and a Kroupa's IMF (dashed-dotted line).
 The best fit Dartmouth model is also reported for comparison (continuous line).
 }
 \label{teotot}
\end{figure}
%%%%%%%%%%%%%%%%%%%%%%%%%%%%%%%%%%%%%%%%%%%%%%%%%%%%%%%%%%%%%%%%%%%%%%%

\citet{paust10} conclude that the global present-day MF for all the GCs in their sample is well matched by a single power-law, in the range of masses between $\sim 0.8~M_{\sun}$ to $0.2-0.3~M_{\sun}$.
While it is likely that a turnover in the MF occurs around these masses \citep{guido,bastian}, our data barely reaches $m=0.5~M_{\sun}$, hence a comparison with single power-law IMF models seems fully adequate. Moreover, \citet{guido} conclude that clusters should have born with very low values of the MF turnover mass ($m_c\simeq 0.15~M_{\sun}$) and it was the subsequent dynamical evolution that drove the drift of $m_c$ to higher values. Since dynamical evolution should be especially slow (or virtually absent) in NGC~2419 it is likely that a single power-law IMF is a good model for this cluster down to very low masses. Indeed, using Eq.~2 by \cite{guido} to obtain rough estimate of $m_c$  we get $m_c\simeq 0.15~M_{\sun}$ for NGC~2419.   

In the upper panel of Fig.~\ref{teotot} we compare the observed LF, in the radial range where it reaches the faintest magnitudes, with four theoretical LFs from the Dartmouth set, with standard values of the physical parameters and different exponent of the underlying power-law IMF. For $M_{F814W}>4.5$, where the slope of the observed LF appears to stabilise, the data points are bracketed between the models with MF slope $x$ between -1 and -2. In the following we will adopt a $x=-1.5$ model as the best-fit. 

%All the theoretical LFs considered in this plot appear to slightly under-predict the data in the %range $2.5\le M_{F814W}\le 4.0$, independently of the adopted $x$. We tried several combinations of %age and chemical composition different from the {\em standard} set and we found that the %discrepancy cannot be alleviated in this way. We conclude that the main reason for this mismatch %lies into the SGB morphology of Dartmouth isochrones, at least in this metallicity and age range. A %detailed inspection of Fig.~\ref{cm} reveals that the region at variable curvature of the %isochrones (well evidenced by the portion of the plot where the three isochrones decouple in each %panel) begins below $M_{F814W}\sim 4.0$ and reaches up to $M_{F814W}\sim 2.0$, while the observed %CMD appears to reach the straight part of the sequence (the base of the RGB) already at $M_{F814W}%%\sim 2.5$. This likely implies a stretch of the star counts over a larger range of magnitudes that %may lead to the observed difference. It is clearly beyond the scope of the present paper to %investigate the origin of this slight mismatch in the CMD morphology, that can be due to some %assumption on the input physics or to the adopted transformations from luminosity + temperature to %magnitude + colour in the given photometric system. We shall see below that this has a negligible %effect on our final scientific goal, since we obtain similar M/L estimate from different LF models, %fitting more or less well this portion of the observed LF.

In the lower panel of Fig.~\ref{teotot} we show that a solar-scaled composition model with the same global metallicity as the {\em standard} model, $[M/H]=-1.8$ \citep[following the prescriptions by][]{scs93}, and age=13.0 Gyr from the Basti set (dashed line) provides an {\em excellent} fit to the observed LF over the whole observed range, adopting  $x=-1.35$, in good agreement with the best-fit slope obtained from the Dartmouth models\footnote{All the Basti models  considered in this paper (either LFs, isochrones or integrated properties) have the Reimers' mass loss parameter $\eta$ equal to 0.4, as expected for a cluster with an extended blue HB. The choice of this parameter has negligible impact on our analysis. The Helium abundance of all the considered models is Y=0.245, independently of the adopted theoretical set.}. Basti models are limited to $m\ge 0.5~M_{\sun}$, that is fully adequate for the range of our LFs but will require some additional assumption to extrapolate the total M/L. We took the very simple approach to extrapolate the Basti model with a spline (dotted line) down to the H-burning limit of the best-fit Dartmouth model, with the additional constraint that at that point the predicted log$N$ must also match. This prescription forces the Basti model to lie below the Dartmouth LF for $M_{F814W}\ga 7.5$, i.e. the regime where stars gives the minimum contribution to the light budget and the maximum contribution to the mass budget. For this reason, the M/L ratio derived from the integration of the Basti LF should be considered as a lower limit. 

The overall shape of the observed LF in the range $2.5\le M_{F814W}\le 4.0$ is also acceptably reproduced by a Padua model with $[M/H]=-1.8$, age=13.5 Gyr and \citet{krou} IMF. On the other hand it appears to overestimate the counts for $M_{F814W}> 4.0$, as expected, since in this range the Kroupa IMF corresponds to a $x=-2.3$ power-law. 
In this case the choice of an arbitrary power-law IMF is not allowed by the web-tool. However we decided to also consider this model, as it is expected to provide an upper limit to the M/L.
The strong peak at $M_{F814W}\simeq 0.0$ is due to the HB that is included in the Padua LFs and not included in those from the other sets of models considered here. 

The best-fit Dartmouth model has been plotted also in this panel, as a reference.
It is interesting to note that all the models correctly predict the position of the RGB Bump and the star counts in this feature \citep[see][and references therein]{f99}, at $M_{F814W}\simeq -1.0$.
It is also worth noting that, according to the thorough analysis by \citet{paust10}, a power-law MF with a slope $-1.5\ga x\ga -1.0$, as observed here, is typical of metal-poor GCs like NGC~2419, and fully compatible with the range of $x$ covered by clusters with similar concentration (see their Fig.~21).  

The three models shown in the lower panel of Fig.~\ref{teotot} will be used in the following to estimate the V band Mass-to-Light ratio ($M/L_V$, always expressed in solar units $[M_{\sun}/L_{V,\sun}$]) of NGC~2419. 
The Dartmouth LF, that provides both a good fit to the observations and a self-consistent model down to the H-burning limit will be the reference for our best-fit estimate. 
The other two models presented in the lower panel of Fig.~\ref{teotot} will be used for an independent  sanity check of our results. They have been explicitly chosen to bracket the reference model. Moreover, they differ in details of the input physics and theoretical/empirical assumptions. Hence they also serve to test the robustness of the derived $M/L_V$ to all these factors.

\subsection{The stellar mass-to-light ratio}
\label{subml}

The integration of the LF and MF of the best-fit models leaves out of the M/L budget two ingredients that may have a significant impact on the the final estimate:

\begin{enumerate}

\item The contribution of the HB and AGB stars, that is not included in the Dartmouth and Basti LFs. 

\item The contribution of remnants of the stellar evolution, i.e. white dwarfs (WD), neutron stars (NS) and black holes (BH).

\end{enumerate}

Concerning the first issue, we have adopted different approaches for the different models, again to explore the effects of different sources of uncertainty.
To include the contribution of HB+AGB in the M/L estimate from the Dartmouth model
we selected all the observed HB and AGB stars using the selection box shown in the right panel of Fig.~\ref{cmsel}, we converted their magnitudes from F606W to V using the transformations presented in Fig.~\ref{trabel}, and, finally, to solar luminosities by assuming $M_{V,\sun}=4.83$  \citep[see][]{maraston}. Then we produced a synthetic HB population with the dedicated Dartmouth web tool\footnote{http://stellar.dartmouth.edu/~models/shb.html} roughly reproducing the observed morphology, and we derived a mean relation to convert V magnitudes into masses. Then summing the luminosity and the mass of all the HB+AGB stars we get the contributions to be added to the corresponding values obtained from the integration of the LF+MF models. It is worth to stress here that HB and AGB stars, that may give raise to 10 to 20 per cent of a GC V light (depending on age, metallicity, IMF and HB morphology of the population), provide only a negligible contribution to the mass budget of the population. Therefore, the final results are completely insensitive to the details of the conversion of their stellar magnitudes into masses.
To make the HB+AGB contribution consistent with the integration of the MS+RGB LF models we (a) normalise the LF models to the observed LF within the half-light radius before performing the LF/MF numerical integration, and (b) we only consider AGB and HB stars lying in the same region. This gives the additional advantage that multiplying the derived luminosity by two we can obtain an estimate of the total cluster luminosity, and consequently $M_V$, that can be compared with the fully independent estimates available in the literature \citep[see Table~\ref{tabpar} and][]{mic}. 
%This way of taking into account the contribution of HB and AGB is clearly an approximation. However %it must be considered that the contribution from these stars to the total $M/L_V$ is not critical: %if we ignore it, the derived $M/L_V$ is increased by 10 per cent. This is in good  agreement with %simple theoretical predictions for simple stellar populations (SSP)\footnote{A simple stellar %population is an ensemble of stars sharing the same age, IMF and chemical composition 
%\citep{rf88}.} from the Evolutionary Flux relation \citep[see][]{rf88}. 

For the Basti model we proceeded as follows. We produced a synthetic CMD with the dedicated web tool, for a simple stellar populations (SSP)\footnote{A simple stellar population is an ensemble of stars sharing the same age, IMF and chemical composition \citep{rf88}.} with age=13 Gyr, [M/H]=-1.8, solar scaled composition and power-law IMF with $x=-1.35$. We use the synthetic population to compute the fraction of the total V light and mass contributed by stars in the HB and AGB phases.
It turns out that these stars provide 18.2 per cent of the total V light and 0.1 per cent of the total stellar mass. The synthetic population has indeed a Blue HB morphology but not so extreme as the observed one. For instance, the faintest (bluest) HB star of the synthetic population has $M_{I}\simeq 1.5$, while the observed HB distribution of NGC~2419 extends down to $M_{I}\simeq 5.5$. Hence the derived fraction should be considered as an upper limit to the light contribution of these stars, that is appropriate since, in the present context, the Basti model has been chosen to provide a lower limit to $M/L_V$.

Finally, it should be noted that for the Padua models the implicit assumption is that all the HB stars lie in a red clump, very different from the actual HB morphology of the cluster. However also this approximation should have only a minor impact on the final $M/L_V$ ratio since, as said, 
the overall contribution of HB stars to the light budget is $\le 20$ per cent, hence, once the number of HB stars is fixed, variations in their luminosity distribution should induce variations in the overall contribution significantly smaller than this, i.e. of order of a few per cent.

Turning to point (ii), the remnants contribute only to the mass budget, since even WDs (the only class of non-dark remnants) are so faint that their contribution to the total light is negligible.
Basti models of the integrated properties of SSPs \citep{perci} provide the mass fraction in remnants obtained by integrating a \citet{krou} IMF with appropriate mass limits and adopting simple initial-final mass relations. 
For a SSP with age= 13~Gyr and $[M/H]=-1.8$, the fraction of the total cluster stellar mass in BH, NS, and WD is 0.03, 0.008, and 0.139, respectively.
All the remnants are produced by processes involving explosions that are generally believed to impart a velocity kick that may lead to the ejection of the object from the cluster \citep[see, e.g.][]{pfahl}. In a recent analysis \citet{krui_remn} estimates that globular clusters retain 0.3-7 per cent of the BH and 0.1-4 per cent of the NS. Combining the mass fractions with the retention factors it should be concluded that the actual contribution of BH and NS to the present-day mass budget of the cluster is $\le$0.2 per cent and $\le$0.03 per cent, respectively. This is clearly negligible and we do not consider it in our computation of $M/L_V$.
For WD we adopt (conservatively) a retention factor of 0.75 \citep[from][]{krui_remn}, so obtaining a total contribution of WD to the present-day mass budget of the cluster of 10 per cent.

Given all the above assumptions, integrating the theoretical V band LF/MF of the Dartmouth model that best fits the observed LF (Fig.~\ref{teotot}) from the RGB Tip down to the Hydrogen-burning limit, and including the contribution of HB+AGB stars and dark remnants as described above, we obtain $M/L_V=$1.49. The associated value of the absolute integrated V magnitude is $M_V=-9.6$, in good agreement with the independent estimates based of surface photometry ($M_V=-9.5$, Table~\ref{tabpar}).
Adopting models with IMF exponent $x=-1.0$ or $x=-2.0$, instead of the best-fit value $x=-1.5$, we obtain $M/L_V=$1.11, and $M/L_V=$2.09, respectively. This gives an idea of the maximum effect on $M/L_V$ that can be obtained by changing the IMF slope, since $x$ values outside the $-2.0<x<-1.0$ range are clearly excluded by the observed LF (at least in the accessible range of magnitudes). 
It is interesting to note that the assumption of $x=-1.0$ or $x=-2.0$ imply total luminosities that are hardly compatible with integrated photometry, i.e. $M_V=-10.1$, and $M_V=-9.0$, respectively. Adopting a 100 per cent retention fraction of {\em all} the dark remnants, the final ratio moves only from $M/L_V=$1.49 to $M/L_V=$1.60.
As a further consistency check we integrated the I band LF/MF of the same model, then converting the total I luminosity into V luminosity by adopting $(V-I)_{0,\sun}=0.688$, from \citet{suncol}, and $(V-I)_{int}=1.05$ as the observed integrated color of NGC~2419, from \citet{peterson}.
With this approach we find $M/L_V=$1.55 for the best-fit model, fully consistent with the result obtained from the V integration. Since the normalization to the observed LF within $r_h$ has been performed in the I (or equivalently F814W) band, this path of integration is expected to provide the most reliable estimate of the total luminosity. We find $M_V=-9.5$, in {\em excellent} agreement with the estimate from integrated photometry. 

The direct V(I) integration of the Basti and Padua models, with the prescriptions described above for remnants and HB+AGB stars, gives V-band mass-to-light ratios of $M/L_V$=1.18(1.24) and $M/L_V$=1.59(1.72), respectively.
The derived values bracket the best-fit estimate from the expected sides (i.e. Basti= lower limit, Padua= upper limit). The absolute integrated magnitude from the integration in I gives $M_V=-9.6$ (Basti) and $M_V=-9.8$ (Padua), both compatible with integrated photometry. 

We conclude that the total stellar $M/L_V$ ratio should lie in the range $1.2\la M/L_V\la 1.7$, with a best-fit value
$M/L_V=1.5$, with a typical 1-$\sigma$ error $\simeq 0.1$, accounting for observational and theoretical uncertainties.

The $M/L_V$ values reported above are lower than the independent estimates obtained by \citet[][$M/L_V=2.05$]{baum09} and \citet[][$M/L_V=1.90$]{iba11a,iba11b} from the kinematics of cluster stars \citep[see also][and references therein]{mcl}, but clearly not incompatible with them\footnote{It should also be considered that the uncertainties in the distance and reddening estimate are not included in our error budget. However these are not expected to have a major impact on our  estimates, since the proper match between observed and predicted LFs is also constrained by distance-independent and reddening-independent features, like the difference in magnitude between the RGB bump and the rapid rise of star counts before the TO point.}.
We defer the detailed discussion of a possible tension between the stellar and the dynamical M/L to a future contribution, dedicated to the analysis of dynamical models of the cluster including dark matter (Nipoti et al., in preparation). 
On the other hand, the agreement with predictions from SSP models is satisfactory \citep{iba11a}. 

The above results are strongly dependent on the assumption that the slope of the MF from our last measured point at $m\sim 0.5~M_{\sun}$ ($M_{F814W}=6.4$) to the H-burning limit, at $m\sim 0.1~M_{\sun}$ ($M_{F814W}\simeq 11.0$) does not vary. While it is quite likely that a single power-law is a good representation of the cluster MF down to very low masses, it is worth exploring the effects of a turnover at low masses \citep{guido,bastian}. With a very conservative approach, given the discussion of Sect.~\ref{teo}, we consider the (unrealistically extreme) case that the cluster does not contain stars with mass lower than 0.3~$M_{\sun}$. Under this condition, we obtain $M/L_V=$~0.89, 0.85, and 1.14 from the Dartmouth, Basti and Padua models, respectively. Thus, even adopting the most conservative assumptions, a robust lower limit to the stellar mass to light $M/L_V>0.8$ is obtained. 

As a final consideration it may be worth noting that the above estimate neglects the presence of binaries, while it is likely that a significant fraction of these systems should be there \citep[likely $\le 20$ per cent, by analogy with other GCs with similar density, see][and discussion therein]{antobin,milo,iba11a}. At the distance of Galactic globulars, all binaries are unresolved and are necessarily included in LFs as single stars slightly brighter than the primary of the system (up to $-2.5{\rm log}2\simeq -0.75$ mag, in case of equal mass components). This leads to two effects, both concurring to obtain $M/L$ estimates that are {\em lower} than the {\em true} values: (a) low-mass stars move from their natural magnitude bin to the brighter bin including the brightened primary, thus leading to an artificial flattening of the MF \citep{mz00,bastian}, and (b) the unresolved binary appears to have a $M/L$ that is lower than the $M/L$ of each of the individual components.
Hence, neglecting binaries in our analysis does not have any effect on the derived lower limits on $M/L_V$. Some simple simulations\footnote{Performed with synthetic populations including binaries extracted from a flat distribution of mass ratios \citep{fisher}, following the same approach as in \citet{iba11a}.} indicate that the {\em true} $M/L_V$ can be larger than our best estimate by a factor from $\sim 1.05$ to $\sim 1.25$ for binary fractions going from 10 per cent to 50 per cent, i.e. an amount comparable with the overall uncertainty in the $M/L_V$ estimate. This effect should help to reconcile the derived stellar $M/L_V$ values with the dynamical estimates. For example, assuming a binary fraction of 20 per cent, $M/L_V$ is expected to grow from $\sim 1.5$ to $\sim 1.7$.

%However, it is well known that the misinterpretation of unresolved binaries as single stars should %artificially {\em 
%This is larger than the  value that has been used several times to fit the velocity dispersion %curves of GCs in alternative theories of gravitation \citep[$M/L_V=$~1.0, see e.g.][and references %therein]{scarpa}.

%Finally it may be worth to push this approach to the extreme, to estimate the mass-to-light of the %population down to the limit reached by the observed LF. Assuming that the cluster has no star %fainter than $M_{F814W}=6.4$ we obtain $M/L_V=$~0.54, 0.48, and 0.52 from Dartmouth, Basti and %Padua model, respectively.

\section{Summary and Conclusions}
\label{summary}

We have used archival HST-WFC3 data to obtain the deepest Luminosity Function so far for the massive and remote globular cluster NGC~2419. The derived LF was corrected for the effects of incompleteness (and its strong radial variation) and bin migration. Comparing the LFs obtained in different radial ranges, also including the wide-field ground-based dataset by D08, we find that the shape of the LF is the same all over the cluster, at least in the range of luminosity covered by our data. This result strongly supports the conclusion by D08 that there is no mass segregation within the cluster, because of the inefficiency of two-body relaxation in establishing the condition of energy equipartition \citep[see][for a detailed discussion]{baum09}. As a consequence, this supports the validity of the hypothesis of constant stellar mass-to-light ratio adopted by \citet{baum09} and \citet{iba11a,iba11b} in their analysis of the cluster kinematics. 

We have fitted our observed LF with theoretical models from three different sets. In the range covered by our data, a power-law IMF with exponent $x\sim -1.5$ provides a reasonable fit to observations, in agreement with the trends found by \citet{paust10}. Integrating the LF/MF of the best fit models and extrapolating the results down to the H-burning limit, with suitable assumptions to account for the contribution of HB+AGB stars and stellar remnants, we find a V band mass-to-light ratio in the range $1.2\la M/L_V\la 1.7$, with a best-fit estimate $M/L_V=1.5\pm 0.1$, slightly lower than the most recent dynamical estimates \citep{baum09,iba11a,iba11b}, but still compatible with them, when all the sources of uncertainty are taken into account. On the other hand, assuming that the cluster has no star with $m\le 0.3~M_{\sun}$, we obtain a robust and conservative lower limit $M/L_V> 0.8$. 

In this context it is worth mentioning two astrophysical processes that may provide a way to increase the mass fraction of dark remnants, thus helping to reconcile the stellar and the dynamical M/L estimates. First, all the available models for the formation of multiple stellar populations in GCs invoke progenitors with total masses much larger than present-day clusters \citep{annib,decre,bekki}. In particular, within the theoretical framework they adopt to explain the claimed large spread of He abundance in NGC~2419, \citet{dic_mult} conclude that the mass of the cluster at its birth should have been from 15 to 30 times the present-day mass. The resulting deep potential well in the first 50-100~Myr of the cluster life may have helped in retaining a larger fraction of dark remnants  within the inner regions of the proto-cluster. Second, a top-heavy IMF \citep{topheavy} may enhance the contribution of dark remnants to the mass budget
of the cluster, but currently available observational evidence does not seem to support this scenario \citep{bastian}. 

The fact that mass-to-light ratio derived from SSP models are typically larger than those found from dynamical analysis is likely accounted for by the selective loss of low-mass stars due to the dynamical evolution of clusters \citep{krui_ml}. The present study suggests that the comparison can be pushed to an higher level, since, on one hand dynamical estimates not simply based on the central velocity dispersion are becoming available \citep[see, e.g.][]{lane11}, and on the other hand there is the observational material \citep{paust10,guido} to derive very robust stellar mass-to-light ratios by integrating observed LF/MF nearly reaching the H-burning limit. These should reflect the present-day stellar content of clusters, not just an educated guess based on average properties of idealized SSP models. These measures based on actual star counts would provide much more stringent (and less model-dependent) constraints on the stellar M/L ratio, to be compared with dynamical estimates. 
%Discrepancies with the results of detailed dynamical models may hint to the presence of Dark %Matter; a good general agreement would provide a nearly conclusive case for the lack of any unseen %non-stellar component in the mass budget of globular clusters.

\section*{Acknowledgments}

We are grateful to an anonymous referee for useful comments and suggestions.
M.B. acknowledge the financial support of INAF through the PRIN-INAF
2009 grant assigned to the project {\em Formation and evolution of massive star
clusters}, P.I.: R. Gratton. 
A.S. acknowledge the support of INAF through the 2010 postdoctoral fellowship grant.
R.I. gratefully acknowledges support from the Agence Nationale de la Recherche though the grant POMMME (ANR 09-BLAN-0228). 
This research has made use
of NASA's Astrophysics Data System.

\label{lastpage}

\end{document}